\documentclass[fleqn,usenatbib]{mnras}

\usepackage{amsmath}
\usepackage{graphicx}
\usepackage{epsfig}
\usepackage{natbib}
\usepackage{longtable}
\usepackage{txfonts}
\usepackage{color}
\usepackage[T1]{fontenc}
\usepackage{microtype}
\usepackage{ulem}   

\usepackage{gensymb}
\usepackage{tabularx}

\newcommand {\be}{\begin {equation}}
\newcommand {\ee}{\end {equation}}
\newcommand {\beq}{\begin {eqnarray}}
\newcommand {\eeq}{\end {eqnarray}}

\newcommand{\powten}[1]{\times 10^{#1}}

\title[Magnetic field structure of compact objects]{Constraints on the magnetic field structure in accreting compact objects from aperiodic variability}

\author[J. M\"onkk\"onen et al.]{
    Juhani M\"onkk\"onen,$^1$\thanks{E-mail: juhemo@utu.fi} 
    Sergey S. Tsygankov,$^{1,2}$
    Alexander A. Mushtukov,$^{3,4}$\newauthor
    Victor Doroshenko,$^{5}$
    Valery F. Suleimanov$^{5}$
    and Juri Poutanen$^{1,2}$
   \\
    $^1$Department of Physics and Astronomy, FI-20014 University of Turku, Finland \\
    $^2$Space Research Institute of the Russian Academy of Sciences, Profsoyuznaya Str. 84/32, Moscow 117997, Russia\\
    $^3$Astrophysics, Department of Physics, University of Oxford, Denys Wilkinson Building, Keble Road, Oxford OX1 3RH, UK\\
    $^4$Leiden Observatory, Leiden University, NL-2300RA Leiden, the Netherlands\\
    $^5$Institut f\"ur Astronomie und Astrophysik, University of T\"ubingen, Sand 1, 72076 T\"ubingen, Germany\\
    }
          
\date{Accepted XXX. Received YYY; in original form ZZZ}
   
\pubyear{2022}

\begin{document}
\label{firstpage}
\pagerange{\pageref{firstpage}--\pageref{lastpage}}
\maketitle
   
\begin{abstract}
We investigate the aperiodic variability for a relatively large sample of accreting neutron stars and intermediate polars, focusing on the properties of the characteristic break commonly observed in power spectra of accreting objects. In particular, we investigate the relation of the break frequency and the magnetic field strength, both of which are connected to the size of the magnetosphere. We find that for the majority of objects in our sample the measured break frequency values indeed agree with estimated inner radii of the accretion disc, which allows to use observed break frequencies to independently assess the magnetic field strength and structure in accreting compact objects.
As a special case, we focus on Hercules X-1 which is a persistent, medium-luminosity X-ray pulsar accreting from its low-mass companion. In the literature, it has been suggested that the complex pulse profiles, the spin-up behaviour and the luminosity-correlation of the cyclotron energy seen in Her X-1 can be explained with a complex magnetic field structure of the neutron star. Here, we connect the measured break frequency to the magnetospheric radius and show that the magnetic field strength derived assuming a dipole configuration is nearly an order of magnitude smaller than the magnetic field strength corresponding to the cyclotron energy. Accordingly, this discrepancy can be explained with the magnetic field having strong multipole components. The multipolar structure would also increase the accreting area on the neutron star surface, explaining why the critical luminosity for accretion column formation is puzzlingly high in this source.
\end{abstract}


\section{Introduction} \label{sec:intro}

In X-ray pulsars (XRPs), the accretion flow from a companion star is disrupted by the magnetic field when the magnetic pressure becomes stronger than the gas ram pressure at a specific distance from the neutron star (NS, see e.g. \citealt{2022arXiv220414185M}). Within this distance, known as the magnetospheric radius, matter flows along the magnetic field lines onto the magnetic poles on the NS surface \citep{King2002}. The accreted matter radiates in X-rays close to the NS and gives rise to the phenomena of XRPs. The observed X-ray flux can be taken to reflect the polar mass accretion rate, because an accretion disc truncated far away from the NS has a negligible contribution to the X-ray flux, especially at high photon energies \citep{1972A&A....21....1P}. At low mass accretion rates, the matter settles onto the NS surface, while above a critical luminosity, the X-ray radiation pressure is able to stop the infalling matter above the hotspots and an accretion column is formed with the magnetic field strength being one of the key parameters determining its structure \citep{1976MNRAS.175..395B,Wang1981, Becker2012, 2015MNRAS.454.2539M}. 

The only direct way of determining the NS magnetic field strength is the cyclotron resonant scattering feature (CRSF) seen in energy spectrum. This cyclotron line is produced by the interaction of X-ray photons with electrons in the strong local magnetic field near the NS surface \citep[see][for a detailed review]{Staubert2019}. The observed centroid energy of the CRSF depends on the magnetic field as
\be
\label{eq:E_cycl}
E_{\rm cycl}= 11.6 B_{12}/(1+z)\ \rm{keV},
\ee
where $z$ is the gravitational redshift ($\sim0.3$ for canonical NS parameters) and $B$ is the magnetic field strength in the line-forming region. In this paper, we use the subscript notation for quantities such that $Q= 10^xQ_x$.

In many sources where the CRSF is observed, the centroid energy of the feature shows a luminosity-dependence. For sources accreting at relatively low luminosities, the correlation is positive \citep{2007A&A...465L..25S,Klochkov2012}, while at higher luminosities, the correlation is observed to be negative \citep{2006MNRAS.371...19T,2007AstL...33..368T,Tsygankov2010,Doroshenko2017,2018A&A...610A..88V}. The critical luminosity dividing these two regimes has been discovered in V~0332+53 \citep{ 2017MNRAS.466.2143D,2018A&A...610A..88V} and 1A~0535+63 \citep{2021ApJ...917L..38K}. In both cases, however, the CRSF energy evolved not only with luminosity but also with time, likely due to complex evolution of the accretion region geometry related to processes at outer magnetosphere. Importantly, these complications suggest that while the cyclotron lines are indeed the only direct way to measure the field in the line-forming region, the relation of measured field to global field of the NS is not trivial, and observed properties of CRSFs can be used to probe it. It also re-emphasises the role of various other indirect methods of field estimation since interpreting the observed evolution of CRSFs can be challenging without this extra bit of information.

One of such indirect methods of determining the field strength exploits the observed aperiodic variability of accreting compact objects. The light curves of accreting compact objects, in general, exhibit variability on timescales from days down to fractions of a second (unlike non-accreting sources, see e.g. \cite{2020A&A...643A.173D} for discussion). The variability has a red noise-like power spectrum and is successfully explained by the perturbation propagation model with viscous processes of the accretion disc. The viscous processes stochastically perturb the local mass accretion rate with a characteristic timescale at each radius and these perturbations then propagate towards the NS \citep{Lyubarskii1997, Kotov2001, Churazov2001, Arevalo2006, Mushtukov2018}. Consequently, the faster variability generated in the inner disc is superposed on the slower variability from the outer disc, which accounts for the observed red-noise in power density spectra (PDSs) of accreting objects.

\cite{Hoshino1993} noted that for magnetised objects, however, there is also a break in the overall power-law continuum spectrum, with a break frequency close to the pulse frequency. This break can be naturally explained by suppression of the variability within the magnetosphere. Indeed, the accretion disc extends down to the magnetospheric radius, inside which matter is guided onto the NS magnetic poles where the X-ray radiation is released. The X-ray flux is modulated by variations in the local mass accretion rate, carrying an imprint of the disc processes. The power-law break seen in the PDS signifies a lack of power at highest variability timescales and is thus connected to the magnetospheric truncation of the disc. We would like to emphasise that unlike in black hole systems, the observed high-frequency variability from the inner regions of a truncated accretion disc has not been strongly suppressed by viscous processes \citep{2019MNRAS.486.4061M}. Therefore, the break frequency is still directly related to the variability timescale at the inner radius, allowing the disc-magnetospheric interface to be studied.

The connection between the break frequency and the accretion disc was exploited by \cite{Revnivtsev2009}, who showed that the break is likely to reflect the Keplerian orbital frequency at the magnetospheric radius in XRPs. Based on this, they suggested a method for estimating the magnetic moment from the break frequency. Similar studies conducted by \citet{Revnivtsev2010}, \citet{Suleimanov2016}, and \citet{2019MNRAS.482.3622S} for intermediate polars, a class of accreting white dwarfs whose accretion disc is truncated within their magnetosphere, similarly to XRPs. Here, we re-visit and expand these works and attempt to connect the break frequencies of sources spanning an unprecedentedly large range of magnetic moments from accreting millisecond pulsars (AMPs) on the low end to intermediate polars on the very high end of this spectrum. 
This allows us to confirm that accretion physics appear to be similar in all cases and form a basic universal rule between the break frequencies and magnetic fields of these objects, and, last but not least, to define and study exceptions to this rule. In particular, we discuss in detail the modelling of the PDS of the well-studied Hercules X-1 (Her~X-1) which appears to deviate from the common trend and discuss reasons for such discrepancy.

The structure of this paper is as follows: we describe the method for estimating the magnetic field strength from the break frequency in Sect.~\ref{sec:break}. We describe data in Sect.~\ref{sec:data} and explain how the break frequencies were derived in Sect.~\ref{sec:results}. We summarise previous research on Her X-1 in Sect.~\ref{sec:herx1} and discuss the implications of our results in this context in Sect.~\ref{sec:discu}.


\begin{table*}
\caption{Key values related to inner radius: $f_{\rm b}$ break frequency, $f_{\rm spin}$ pulse frequency, $B$ magnetic field strength, $\dot{M}$ mass accretion rate and $d$ distance. The uncited pulse frequencies are the values of \textit{Fermi} Gamma-Ray Burst Monitor Pulsar Project except for Her X-1 for which the value was determined in this work. Distances for pulsars are from \protect\cite{2021MNRAS.502.5455A} summarising \protect\cite{BailerJones2018} unless otherwise cited. Break frequencies marked with an asterisk were measured from a PDS averaged from several observations (corresponding mass accretion rate being averaged as well). Values for intermediate polars were measured by \protect\cite{2019MNRAS.482.3622S} unless a citation is given.}
\label{tab:pulsars}
\begin{tabular}{lllllll}
\hline
Pulsar & $f_{\rm b}$ [Hz] & $f_{\rm spin}$ [Hz] & $B$ [$10^{12}$ G] & $\mu$ [$10^{30}$ G cm$^3$] & $\dot{M}$ [$10^{16}$ g s$^{-1}$] & $d$ [kpc]  \\
\hline

1A 0535+262     & $0.14\pm 0.02$ & 0.00969 & $5.1\pm 0.1 ^{1}$ & $8.8\pm 0.1$ & $56.8\pm0.7$ & $2.1\pm0.3$  \\

1A 1118$-$616   & $0.28\pm 0.03$ & 0.00250 & $6.2 ^{2}$ & $10.7\pm 0.1$ & $7.4\pm0.2$ & $2.9\pm0.3$  \\

4U 0115+63      & $0.29\pm 0.02$ & 0.277 & $1.3\pm 0.1 ^{3}$ & $2.2\pm 0.1$ & $35\pm3$ & $7\pm 1^4$ \\

Cep X-4         & $0.030\pm0.008$ & 0.0151 & 3.4$^5$ & $5.4\pm 0.1$ & $19.5\pm0.5$ & $10\pm2^6$ \\

GRO J1008$-$57  & $0.25\pm 0.02$* & 0.0107 & $8.7\pm0.3 ^{7}$ & $15.0\pm 0.3$ & $10\pm1$ &  $3.6\pm 0.5^{8}$  \\

GRO J1744$-$28  & $100\pm 10$ & 2.14 & $0.50\pm 0.02 ^{9,10}$  & $0.86\pm 0.02$ & $49\pm5$ & 8$^{11}$  \\

GX 304$-$1        & $0.41\pm 0.04$ & 0.00364 & $6.1\pm 0.1 ^{12}$ & $10.5\pm 0.1$ & $11.6\pm0.2$ & $2.0\pm0.2$ \\

Her X-1         & $6.5\pm 0.2$ & 0.808 & $4.5\pm 0.1 ^{13}$ & $7.8\pm 0.2$ & $19.7\pm0.3$  & $5.0\pm0.7$  \\

MXB 0656$-$072 & $1.08\pm0.03$* & 0.00629 & 3.7$^{14}$ & $6.4\pm 0.1$ & $28\pm3$ & $5\pm 2$  \\

V 0332+53       & $0.206\pm 0.006$ & 0.229 & $2.9\pm 0.2 ^{15}$ & $5.0\pm 0.2$ & $25\pm3$ & $5\pm 1^4$ \\

XTE J1946$-$274 & $1.05\pm0.02$* & 0.0635 & 4$^{16}$ & $6.9\pm 0.1$ & $90\pm10$ & $13\pm4$  \\

\hline

AMP & $f_{\rm b}$ [Hz] & $f_{\rm spin}$ [Hz] & $B$ [$10^{8}$ G] & $\mu$ [$10^{30}$ G cm$^3$] & $\dot{M}$ [$10^{16}$ g s$^{-1}$] & $d$ [kpc] \\

\hline

IGR J17480$-$2446 & $851\pm 4 ^{17}$ & 11.0$^{18}$ & $30\pm10^{19}$ & $0.0015\pm 0.0003$ & $59\pm 3 ^{20}$ & $6\pm 2 ^{11}$  \\

IGR J17511$-$3057 & $250\pm 20 ^{21}$ & 245$^{22}$ & $<3.5^{23}$ & $<0.0004$ & $4.7\pm 0.3 ^{21,24}$ & $5\pm 1^{25,26}$  \\

SAX 1808.4$-$3658 & $700\pm 10 ^{27}$ &  401$^{28}$ & $0.8\pm0.5^{29}$ & $0.0003\pm 0.0002$ & $1.9 ^{30}$ & $3.5\pm 0.1 ^{31}$  \\

\hline

Intermediate polar & $f_{\rm b}$ [Hz] & $f_{\rm spin}$ [Hz] & $B$ [MG] & $\mu$ [$10^{30}$ G cm$^3$] & $\dot{M}$ [$10^{16}$ g~s$^{-1}$] & $d$ [pc] \\

\hline

EX Hya & 0.013 & 0.000249 & $0.029$ & $13.75\pm 0.03$ & $(13.0\pm0.6)\times 10^{-2}$ & $57.0\pm0.2$  \\

GK Per & 0.0017 & 0.00285 & $0.5^{32}$ & $180\pm 40$ & $79\pm3$ & $442\pm9$  \\

NY Lup & 0.0005 & 0.00144 & $>4^{33}$ & $>500$ & $21\pm2$ & $1270\pm50$  \\

\hline
\end{tabular}
\centering{\\
$^{1}$\protect\cite{Terada2006},
$^{2}$\protect\cite{Doroshenko2010},
$^{3}$\protect\cite{2007AstL...33..368T},
$^{4}$\protect\cite{2019A&A...630A.105R},
$^{5}$\protect\cite{Mihara1991},
$^6$\protect\cite{Treuz2018, BailerJones2018},
$^{7}$\protect\cite{Bellm2014},
$^{8}$\protect\cite{Riquelme2012},
$^{9}$\protect\cite{Doroshenko2015}
$^{10}$\protect\cite{DAi2015}
$^{11}$\protect\cite{Nishiuchi1999}
$^{12}$\protect\cite{Yamamoto2011}, 
$^{13}$\protect\cite{Staubert2016}, 
$^{14}$\protect\cite{McBride2006},
$^{15}$\protect\cite{2006MNRAS.371...19T},
$^{16}$\protect\cite{Doroshenko2017},
$^{17}$\protect\cite{Altamirano2012}, 
$^{18}$\protect\cite{2010ATel.2929....1S},
$^{19}$\protect\cite{Miller2011}, 
$^{20}$\protect\cite{Linares2012}, 
$^{21}$\protect\cite{Kalamkar2011}, 
$^{22}$\protect\cite{2009ATel.2197....1M},
$^{23}$\protect\cite{Papitto2016},  
$^{24}$\protect\cite{Papitto2010}, 
$^{25}$\protect\cite{Altamirano2010},
$^{26}$\protect\cite{Ibragimov2011},
$^{27}$\protect\cite{Bult2015}, 
$^{28}$\protect\cite{1998Natur.394..344W},
$^{29}$\protect\cite{Ibragimov2009}, 
$^{30}$\protect\cite{DiSalvo2019}, 
$^{31}$\protect\cite{Galloway2006},
$^{32}$\protect\cite{2018MNRAS.474.1564W},
$^{33}$\protect\cite{2012MNRAS.420.2596P}.
}
\end{table*}

\section{Relation of the magnetic field and break frequency}
\label{sec:break}

According to the perturbation propagation model, the X-ray emission released over the magnetic poles of a NS or a white dwarf is modulated by the mass accretion rate variability accumulated over the radial scale of the accretion disc \citep{Lyubarskii1997}. In the PDS of the X-ray light curve, the resulting aperiodic variability follows a power law until breaking to a steeper slope due to the lack of noise generated inside the magnetosphere. The power-law break occurs at the frequency which corresponds to the variability timescale at the innermost radius of the accretion disc. Therefore, \cite{Revnivtsev2009} suggested that the break frequency, taken to be the Keplerian frequency at the inner disc radius, can be used for estimating the magnetospheric radius and, consequently, the magnetic field strength. 

The magnetospheric radius of the accretion disc, that is, the inner disc radius, can be derived from the Alfv\'en radius $R_{\rm A}$ which is estimated by equating the magnetic pressure of a pure dipole field with the gas ram pressure
\citep{Lamb1973,King2002}: 
\begin{equation}
\label{eq:r_m}
    R_{\rm m} \simeq \Lambda R_{\rm A}\simeq 5.1\times 10^8 \Lambda m^{-1/7} \dot{M}_{16}^{-2/7} \mu_{30}^{4/7}\ \rm{cm,}
\end{equation}
where $\Lambda$ is the parameter describing the geometry of the accretion flow, $m=M_*/M_\odot$ is the mass of the compact object in units of solar masses, $R_{*}$ is the radius of the compact object, $\dot{M}$ is the mass accretion rate, $\mu_{30}=B_{12}R_{*,6}^3$ is the dipole magnetic moment, and $B_{12}$ is magnetic field strength at the magnetic pole of a NS. We take $\Lambda=0.5$, around which the values found in theoretical models and simulations of accretion disc typically are \cite[see e.g.][]{Ghosh1979}. In this work, we also assume $m=1.4$ and $R_{*,6}=1.2$ for NSs \citep{Suleimanov2017}.
Masses and radii for white dwarfs were taken from \citet{2019MNRAS.482.3622S}.

Regarding variability timescales, we note that matter in the accretion disc has a Keplerian orbital velocity around the compact object. This corresponds to the Keplerian frequency 
\begin{equation}
\label{eq:f_K}
    f_{\rm K}(R)= \Omega_{\rm K}(R)/2\pi = \frac{1}{2\pi}\sqrt{\frac{GM_*}{R^3}} = 1.83\sqrt{\frac{m}{R_8^3}} \,\,  \rm{Hz,} 
\end{equation}
where $R$ is the distance from the compact object. For our study of the magnetospheric boundary, we want to know the maximal frequency of initial variability generated in the accretion disc. Incidentally, this is the Keplerian frequency at the magnetospheric radius, $f_{\rm K,m} = f_{\rm K}(R_{\rm m})$. Combining equations \eqref{eq:f_K} and \eqref{eq:r_m} we get 
\begin{equation}
    f_{\rm K,m}\simeq 0.1\ \Lambda^{-3/2} m^{5/7} \dot{M}_{16}^{3/7} \mu_{30}^{-6/7} \,\,{\rm Hz.}
\end{equation} 
Assuming that the break frequency is related to this Keplerian frequency, these represent its basic dependencies on the key physical quantities. Because we are especially interested in the magnetic moment, it is useful to examine a break frequency that has been normalised with respect to the mass of the compact object and the mass accretion rate:
\begin{equation}
    \label{eq:f_b_norm}
    f_{\rm b,norm} = f_{\rm b}\, \left( \frac{M_*}{1.4 M_\odot}\right)^{-5/7} \left(\frac{\dot{M}}{6.46\times 10^{16}\ {\rm g\ s}^{-1}}\right)^{-3/7},
\end{equation}
in which $6.46\times 10^{16}$~g~s$^{-1}$ corresponds to the luminosity of $10^{37}$~erg~s$^{-1}$ in NSs. The normalised break frequency depends only on the magnetic moment as $f_{\rm b,norm}\propto \mu_{30}^{-6/7}$.

We adopt these dependencies on the physical parameters, motivated by \cite{Revnivtsev2009} who confirmed that $f_{\rm b}\propto \dot{M}^{3/7}$ in agreement with the assumptions of a thin disc and Ghosh \& Lamb torque balancing at the magnetosphere. Furthermore, the simulations by e.g. \cite{2003ApJ...595.1009R,2012MNRAS.421...63R} indicate that the magnetospheric radius is determined by the balancing of the magnetic pressure with the gas ram pressure. There have been alternate developments in determining the border of the magnetosphere by, for example, \cite{1996ApJ...469..765L}, \cite{1999ApJ...527..910L} and \cite{2012MNRAS.420.1034C} who considered the limits set by the accretion flow in the magnetosphere. Thus, the exact form of the magnetospheric radius may thus differ from the one introduced above. Possible sources of error are further discussed in section~\ref{sec:discu}.

Although the Keplerian motion of disc matter serves as the basis for the variability generated in the disc, the break frequency itself may not correspond to the $f_{\rm K,m}$.
To estimate the probable range of frequencies for the break, we must consider other variability timescales in the disc. Namely, it is likely that the mass accretion rate fluctuates on the dynamo timescale which is related to the magnetic alignment timescale \citep{2004MNRAS.348..111K}. Thus, the break frequency could be expected to correspond to the dynamo frequency $f_{\rm d}$ at the innermost radius, the local initial timescale of variability \citep{2019MNRAS.486.4061M}. The dynamo frequency is directly proportional to the Keplerian frequency but smaller by a factor of a few, so a coefficient $k_{\rm d}$ connects the two frequencies as $f_{\rm d}= f_{\rm K} / k_{\rm d}$. The exact value of the coefficient depends on the details of the magnetorotational instability: for example, observational results of \cite{Doroshenko2014} give $k_{\rm d}\approx 2$ which is in line with earlier work of \cite{Ghosh1979} on the disc-magnetospheric interface.
Thus for XRPs, the Keplerian frequency provides an upper limit for the break frequency, and as a lower limit we take the dynamo frequency corresponding to factor $k_{\rm d}=10$ given in \cite{2004MNRAS.348..111K}. The dependencies on physical quantities are the same in any case and the $f_{\rm b,norm}$ can be used for the study. A large sample of XRPs with different properties is needed to investigate the characteristic frequency of the initial variability in a statistically meaningful way.

The break frequency has a relatively weak dependence on all parameters, and therefore it is important to investigate it for a wide range of magnetic moments (the situation with mass accretion rate is somewhat easier as there are objects varying in luminosity by more than five orders of magnitude, \citealt{2020MNRAS.491.1857D}). Such investigation implies that a very heterogeneous sample of objects has to be considered, but fortunately, $f_{\rm b,norm}$ can be compared between both X-ray pulsars and intermediate polars. We describe next what objects were chosen for the analysis and how the break frequency was measured from their PDSs.
 

\section{Data reduction and analysis} 
\label{sec:data} 

In our study, we combined parameter values reported in the literature as well as values measured in our own analysis. Specifically, we chose 11 highly magnetised XRPs, 3 AMPs and 3 intermediate polars whose parameters are summarised in Table \ref{tab:pulsars}. 

\subsection{X-ray pulsars}

For highly magnetised accreting NSs we determined the break frequencies using data from the Proportional Counter Array (PCA) instrument onboard \textit{Rossi X-ray Timing Explorer} (\textit{RXTE})  because it has very good temporal resolution and a large effective area essential for timing analysis on shorter timescales. Programs from \textsc{Heasoft} software package\footnote{\url{https://heasarc.gsfc.nasa.gov/lheasoft/}} v. 6.29 were used for the data reduction and analysis. Light curves and spectra were constructed following standard procedures of \textit{RXTE} Cook Book guide\footnote{\url{https://heasarc.gsfc.nasa.gov/docs/xte/recipes/cook\_book.html}}. The pulse frequencies given in Table \ref{tab:pulsars} are the latest values from \textit{Fermi} Gamma-Ray Burst Monitor Pulsar Project\footnote{\url{https://gammaray.msfc.nasa.gov/gbm/science/pulsars.html}} taken at the time of analysis (December 2021).

\textsc{Powspec} was used to construct the PDS from light curves for a frequency range between 0.5~mHz and 1024~Hz. For Her~X-1, a constant number of 20 PDS points were binned linearly in order to use the Whittle likelihood in fitting. The number of points binned was chosen such that the PDS became smoothed but a considerable number of points remained at low frequencies for fitting. The power was normalised with Miyamoto normalisation \citep{1991ApJ...383..784M}. For other XRPs, the binning was varied depending on the statistics. To increase the fitting accuracy, the PDS for GRO J1008$-$57, MXB 0656$-$072 and XTE J1946+274 were constructed from five observations at the peaks of their outbursts. When possible, we sought uniformity by measuring the break frequency at a luminosity close to the determined peak luminosity of Her X-1, $L_{\rm peak}= (3.05\pm0.05)\times 10^{37}$~erg~s$^{-1}$. 

The luminosities were calculated based on the distances to the sources reported in literature \citep{Treuz2018,BailerJones2018, 2019A&A...630A.105R,2021MNRAS.502.5455A} and listed in Table~\ref{tab:pulsars}, and fluxes derived from spectral analysis in this work for each observation. In particular, Standard2 spectra were modelled in the energy range of 3--50~keV. The model consisted of a photo-absorbed cut-off power-law with an additive Gaussian component for the iron fluorescence feature around 6.5~keV and a multiplicative Gaussian absorption feature for the CRSF when needed: \textsc{phabs*gabs*(powerlaw*highecut+gauss)}. Bolometric luminosity was estimated from the unabsorbed model flux in an energy range from 0.5 to 100~keV. This was transformed into mass accretion rate by $\dot{M}=LR_*/(GM_*)$. We emphasise that the sole goal of the spectral analysis was to estimate fluxes, so in spite of including the CRSF in the spectral model, we used cyclotron energies from the dedicated studies in the literature to increase the robustness and accuracy of the results.

\subsection{Her X-1}

Her X-1 is a persistent XRP accreting from its binary companion HZ~Her through an accretion disc which is thought to be tilted and warped. The system is seen almost edge-on so the occultation by the precessing accretion disc causes a 35~d superorbital periodicity to the apparent X-ray luminosity with alternating on-states and off-states \citep{1976ApJ...209..562G}. During the 35~d cycle, the view to the NS is the most unobscured at the peak of the bright main-on phase \citep{1991A&A...245L..29B}. We examined a representative main-on phase around MJD 50710 and deemed it the most reliable to measure the PDS at the peak of the main-on when the source is least obscured. Additionally, we excluded observations when the source was eclipsed by its companion or the accretion stream \citep{Schandl1996}. As a sign of stability, we note that another well-covered main-on around MJD 52600 had similar apparent luminosity and peak PDS as around MJD 50710. The distance to the source was taken to be $5.0_{-0.7}^{+0.6}$~kpc \citep{BailerJones2018,2021MNRAS.502.5455A}.

The measurement of break frequency for Her X-1 is complicated by the fact that the source has a complex pulse profile. In the PDS, the pulse profile is seen as an extensive "forest" of pulse harmonics and the break frequency lies within them. Concerns on retrieving the exact break frequency \citep[see][]{1997ApJ...476..267L} lead us to create a model which convolves the PDS of folded regular pulsations with the assumed underlying broken power law. The model and fitting procedures are described in detail in Appendix \ref{sec:convolution}. 

\subsection{Accreting millisecond pulsars}

To include sources with low magnetic fields we chose examples from the literature. The PDS of these objects appear to be governed by a number of distinct Lorentzian features \citep[see an example in e.g. ][]{Altamirano2012} so the determination of the break is not very straight-forward. We interpreted the upper kHz quasi-periodic oscillation (QPO) as a sign of the fastest variability of the accretion disc at the innermost radius \citep{vanderKlis2000,Bult2015}. Based on the observed pulsations it can be assumed that the disc is truncated by the NS magnetic field and not by the boundary layer so we can compare the centroid frequency of the upper kHz Lorentzian to the break frequencies of high-$B$ sources. 

For two sources, the magnetic field strength is independently derived from the inner disc radius determined from energy spectral fits: For IGR J17480$-$2446, the magnetic field strength of $(3\pm1)\times 10^9$~G (for a radius of $10^6$~cm) is based on relativistic emission line \citep{Miller2011}. For SAX 1808.4$-$3658, $B=(8\pm5)\times 10^7$~G was derived from the visibility condition of the hotspots contributing to the pulse profile \citep{Ibragimov2009}. In case of IGR J17511$-$3057, the magnetic field strength could be constrained to $B<3.5\times 10^8$~G based on the spin-down rate \citep{Papitto2016}.


\begin{figure}
\includegraphics[width=\columnwidth]{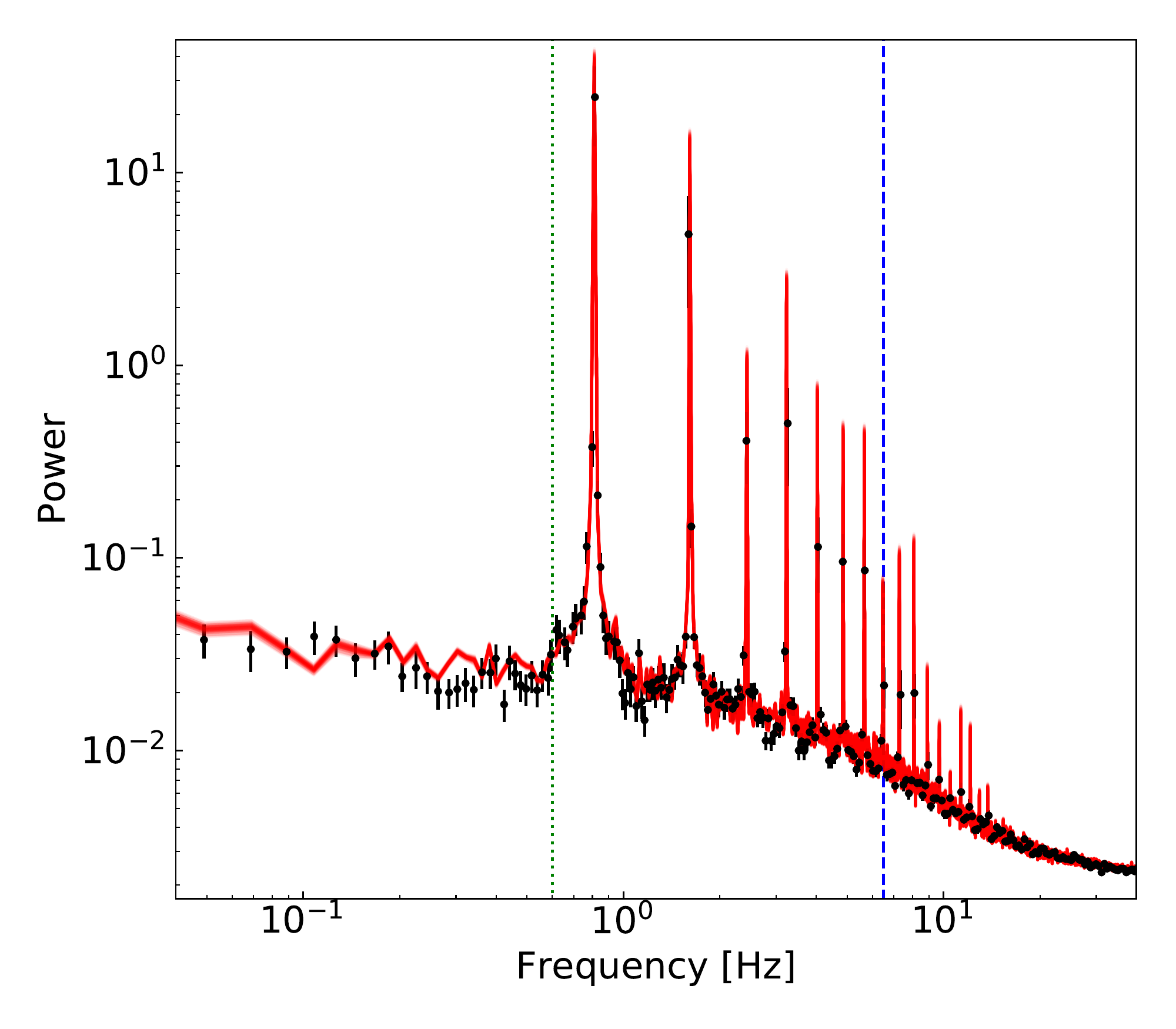}
\caption{Power density spectrum of Her X-1 obtained from a single observation performed during the MJD 50710 main-on at the peak luminosity of $(3.05\pm0.05)\times 10^{37}$~erg~s$^{-1}$. The points were geometrically rebinned with a base of 1.02. The red lines present a hundred models with parameters randomly sampled from the Markov Chain Monte Carlo run. The blue dashed vertical line marks the measured break frequency while the green dotted line shows the break which is expected for the magnetic field estimated from the cyclotron energy and assuming dipole field configuration.}
\label{fig:pds_herx1}
\end{figure}

\subsection{Intermediate polars}

\cite{Revnivtsev2010} studied a type of accreting white dwarfs, intermediate polars, where accretion discs are  truncated by the white dwarf magnetic field. \cite{Revnivtsev2010} showed that the flux variability of intermediate polars in the optical band also follows a broken power law, in which the break can be associated with the disc truncation radius, similarly to highly magnetised XRPs. 
Similar study was conducted by \cite{2019MNRAS.482.3622S} in the X-ray band, where simultaneous fitting of the break frequency and energy spectrum was used to determine the mass and radius as well as the magnetic field strength of several white dwarfs. Therefore, their results proved to be useful for our analysis to expand the sample of sources to magnetic moments larger than those of XRPs. We emphasise that a small scatter of the white dwarf masses deduced by \cite{2019MNRAS.482.3622S} when compared to independent mass estimates suggests that the underlying assumptions are correct and the derived magnetospheric radii are thus reliable.
Moreover, for two sources, the magnetic moment had been independently measured. Polarimetry was used by \cite{2012MNRAS.420.2596P} for NY~Lup to get a lower limit of $B>4$~MG. \cite{2018MNRAS.474.1564W} fit the energy spectra of GK Per in its outburst and quiescent states to determine the magnetic field strength of $5\times 10^5$~G by combining models for the Alfv\'en radius and accretion column temperature. As a third example white dwarf, we chose the outbursting EX~Hya, even though there was no independently measured value for the magnetic moment. The value of B $\approx 2.9 \times 10^4$\,G derived by simultaneous energy and power spectral fits by \cite{2019MNRAS.482.3622S} was nevertheless used to highlight the range of magnetic moments occupied by intermediate polars.
For each of these sources, we used the radius values determined by \cite{2019MNRAS.482.3622S} when calculating the magnetic moment.

\section{Results} \label{sec:results} 

The PDS of Her X-1 shows the typical aperiodic variability modulated by the periodic signal: The continuum is similar to a broken power law with harmonics of the pulsation at frequency $f_{\rm P}=0.81$~Hz superposed. The data are presented in Fig. \ref{fig:pds_herx1} along with a sample of models from the Markov Chain Monte Carlo (MCMC) run. Remarkably, the break frequency is measured at $6.5\pm0.2$~Hz which is an order of magnitude higher than the expected $f_{\rm K}(R_{\rm m})\approx 0.6$~Hz based on the CRSF, even after the convolution with pulsations was taken into account. We note that subtracting pulsations in a manner similar to \cite{Revnivtsev2009} and modelling the PDS with a simple broken power law also leads to comparable results.

The chosen analytic expression for the PDS continuum may introduce systematic error, as is evident in the placement of the break over a pulse harmonic. In any case, even for complexly pulsed Her X-1, fit results for the break frequency remain similar whether we use the convolving model or another model without convolution, such as a smoothly broken power law with additive Lorentzian functions for the strongest pulse peaks. 
Thus, we used a simple, broken power-law model for fitting all the other XRPs from our sample.

If we conservatively take the measured break frequency $f_{\rm b}=6.5\pm 0.2$~Hz for Her X-1 to correspond to the Keplerian frequency at the inner disc radius, we get a radius of $R= (4.9\pm 0.1)\powten{7}$~cm from Eq.~\eqref{eq:f_K}. Compared to the magnetospheric radius calculated for a CRSF-based dipole field, this radius is six times smaller. The discrepancy is even higher, if we are to consider dynamo frequencies. 

\begin{figure}
\includegraphics[width=1\columnwidth]{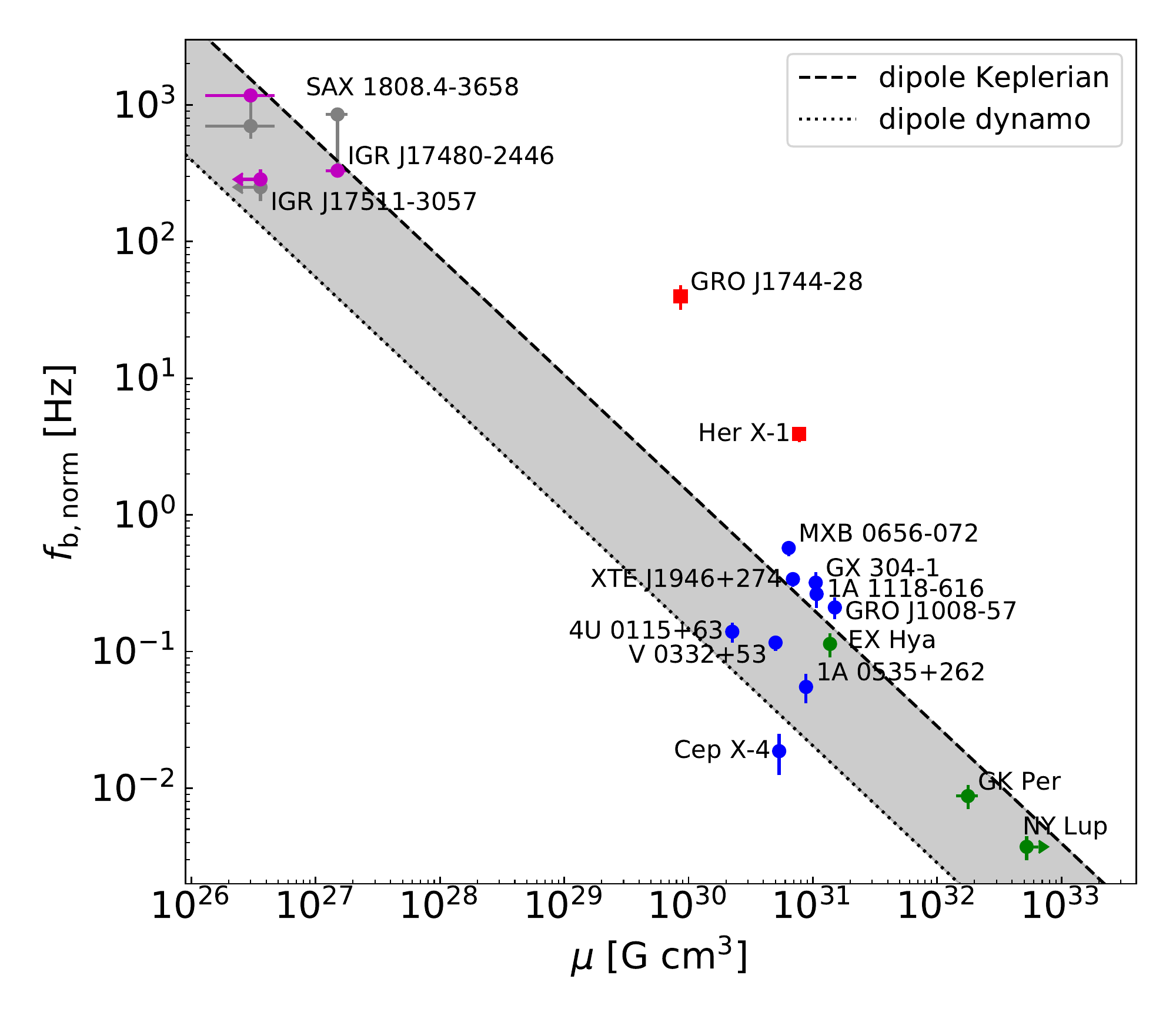} 
\caption{Normalised break frequencies $f_{\rm b,norm}$, as given by equation \eqref{eq:f_b_norm}, of several XRPs, AMPs, and intermediate polars plotted against their magnetic moments $\mu$. Different colours indicate different object types: magenta mark accreting pulsars with kHz quasi-periodic oscillations (gray are the unscaled measurements), blue mark XRPs with a break in PDS and green is for the intermediate polars. Red squares are the outliers discussed in the text. The interval depicts the theoretical prediction for the probable range of break frequencies: The upper limit, the dashed line is the maximal frequency in the disc, the Keplerian frequency of the magnetospheric radius for a pure dipole field with $\Lambda=0.5$.
The lower limit, the dotted line shows the dynamo frequencies estimated to be one tenth of the Keplerian frequency for the dipole field configuration.}
\label{fig:fb_mu}
\end{figure}

This discrepancy becomes even more obvious when compared to other sources. To illustrate this point, we combined measured or collected from the literature break frequency values and field estimates for all the objects in our sample. As discussed in Sect.~\ref{sec:break}, the break frequency is expected to depend on both the magnetic moment of a compact object and the mass accretion rate at a given time as well as the mass of the compact object. Therefore, in Fig.~\ref{fig:fb_mu}, we plotted the normalised break frequency $f_{\rm b,norm}$, given in equation \eqref{eq:f_b_norm}, against the dipolar magnetic moment $\mu$.
In Fig.~\ref{fig:fb_mu}, for sources with low magnetic moments (AMPs), the original QPO frequencies taken as the break are also shown. Uncertainty in distance was not taken into account here, but it is relatively small in most cases considering the weak dependence of break frequency on mass accretion rate.

As seen in Fig.~\ref{fig:fb_mu}, for the wide range of magnetic moments considered, the break frequencies appear well in agreement with $\mu^{-6/7}$. This is in agreement with the expectation that the break frequency is related to the Keplerian timescale at the inner radius of an accretion disc truncated by the compact object's magnetosphere. Most of the sources have a break above the estimated dynamo frequency of $\frac{f_{\rm K}}{10}$ (note, however, that the scaling between the dynamo frequency and break frequency is in general rather uncertain). Although many break frequencies slightly exceed the supposedly maximal Keplerian frequency, the largest deviations are about a factor of two when excluding Her X-1 and GRO~J1744$-$28. These two sources instead have break frequencies more than an order of magnitude higher than expected. In the next section, we summarise studies of the magnetic field structure and the presence of multipole components in Her X-1, and discuss our results in this context in Sect.~\ref{sec:discu}.


\section{The magnetic field of Her X-1}
\label{sec:herx1}

Her X-1 was the first XRP in which a CRSF was detected. Its centroid energy 40~keV corresponds to a surface magnetic field of $B\sim 4.5\times 10^{12}$~G \citep{Truemper1978}. Furthermore, the CRSF is observed to be positively correlated with luminosity, which is typical for sub-critical XRPs \citep{2007A&A...465L..25S,Vasco2011}. However, the critical luminosity value required for Her X-1 in this case is too high in comparison with the theoretical predictions \citep{2015MNRAS.447.1847M}. \cite{Becker2012}, in order to solve this problem, assumed a lower value of the accretion flow geometry parameter $\Lambda=0.1$, corresponding to a much smaller magnetosphere in comparison to a broadly accepted $\Lambda=0.5$.

Several independent studies indicate that the magnetic field of Her X-1 contains strong multipole components. The studies of accretion torque by \cite{Ghosh1979a} and \cite{Wang1987} showed that the period change behaviour in Her X-1 requires a magnetic field strength smaller by a factor of a few with respect to the magnetic field strength determined from the cyclotron line. In addition to that, \cite{Shakura1991} studied the complex pulse profiles and could explain them by including both dipole and quadrupole components to the magnetic field of the NS, which allow for an additional accreting ring to occur around one magnetic pole. This was also in agreement with the pulse phase dependence of the CRSF. \cite{1994A&A...286..497P} and \cite{2013MNRAS.435.1147P} fit pulse profiles with such a model and got results in agreement with the existence of an additional accretion ring. Alternatively, the pulse profile evolution can be explained with a complex emission beam combination and hotspot occultation by the accretion disc with a small inner radius $(2-5)\times 10^7$~cm \citep{2000ApJ...539..392S}. Similarly, \cite{2004MNRAS.348..932L} examined the pulse profiles and noted that an offset of one polar region is required. 

Since the NS magnetic field truncates the accretion disc and dictates the polar accretion, it is likely that we can see the presence of multipole components in the PDS of the X-ray light curve -- not only in the pulsations but also in the aperiodic noise. Therefore, our result of an unexpectedly high break frequency may not be that surprising after all. Next, before concluding, we consider possible reasons for the deviation. 

\section{Discussion and conclusions} 
\label{sec:discu}

As seen in Fig.~\ref{fig:fb_mu}, the break frequencies are all in all anticorrelated with the magnetic moment, forming a fundamental plane across various magnetic accretors. That is, the larger the magnetosphere, the smaller the break frequency. However, there is also a relatively large scatter in their break frequency values, especially for the highly magnetised XRPs. Various effects can cause scatter of break frequencies and deviations from the maximal frequency:
\begin{enumerate}
    \item According to \cite{Bozzo2018}, the inclination of the NS dipole with respect to the accretion disc plane affects the magnetospheric radius by a factor of two between the extreme cases. Such objects were simulated by \cite{2021MNRAS.506..372R} to study the tilting and warping of the disc.
    
    \item The basis of the magnetic field strength measurement, the centroid energy of the CRSF is seen to change with respect to the pulse phase as well as during the outburst evolution \citep[see e.g.][]{2018A&A...610A..88V}.
    
    \item Another source of error may lie in the applied correction of break frequencies given in equation \eqref{eq:f_b_norm}, even though the dependencies is weak. (Note that the scaling is only for comparison and does not take into account any physical constraints nor any relativistic effects.) 
    
\item The inaccuracy of measured distances translates to large errors in the mass accretion rate. On the other hand, it is counter-acted by the weak dependence of the break on mass accretion rate.
        Additionally, the dependence of break frequency may not be exactly $\dot{M}^{3/7}$ \citep[see e.g.][especially for weak magnetic fields]{vanderKlis2001,Kulkarni2013,Bult2015,Kalamkar2011}.
       \cite{Mendez1999} discusses how in some AMPs, luminosity may not be a good measure of the mass accretion rate, making QPO measurements unreliable for estimating the inner radius.
 
\end{enumerate}

Despite the uncertainties, the break frequency acts as a rough measure of the magnetic field. For estimating the magnetic moment, it is good to note that there are alternative theories for the interaction between the accretion disc and the magnetosphere, such as the funnel flow perspective \citep{1996ApJ...469..765L,1999ApJ...527..910L}. Specifically, the inner radius predicted by equation (19) in \cite{1999ApJ...527..910L} is twice larger (assuming the square of Alfv\'en Mach number at the boundary of the accretion curtain and disc $D_{\rm d}=0.5$, velocity of matter at the boundary equal to the sound speed $v_{\rm d}=c_{\rm s}=2.5\times 10^6$~cm~s$^{-1}$ and ratio of the width of the curtain to the distance from the compact object $\Delta \varpi/\varpi=0.1$) than the one given by the radius of equation \eqref{eq:r_m}. The corresponding break frequency would be smaller but still within the current range between Keplerian and dynamo frequencies.
The picture is also complicated by simulations of e.g. \cite{2003ApJ...595.1009R} which suggest that disc matter is magnetically braked already outside the magnetospheric radius. This means that the maximal variability generated in the braked disc is slightly slower than in the non-braked disc, which leads one to overestimate the size of the magnetosphere if only the Keplerian angular velocity is considered. 

In our sample, two XRPs, Her X-1 and GRO J1744$-$28, stand out having a break frequency an order of magnitude higher than expected, beyond a combined effect of all the uncertainties. It is interesting to note that the presence of strong multipole field components has been suggested for both objects in the literature \citep{Shakura1991,2013MNRAS.435.1147P,2019A&A...626A.106M,2020A&A...643A..62D} based on multiple independent considerations. If that is indeed the case, one could expect to observe a reduced effective magnetosphere size (compared to the estimate for a dipole field) which could help to explain the observed high break frequency values.

To illustrate this point we wish to consider how a non-dipolar magnetic field configuration would affect the inner radius of the accretion disc and subsequently, the break frequency, similarly to what was discussed in \cite{2019A&A...626A.106M}. The actual 3-dimensional shape of a multipole magnetic field is complex and may have considerable effects on observational properties because the field lines of such a field cross the NS surface between the poles \citep{Mastrano2013,Das2022}. For a symmetric quadrupole, the field lines trace a ring along the NS equator, allowing an additional accretion channel \citep{Long2007}. Depending on the disc inclination with respect to the NS, this equatorial cleft of a quadrupole field could decrease the inner radius of the accretion disc. \citet{Shakura1991} and \citet{2013MNRAS.435.1147P} suggested an asymmetric composition of both dipole and quadrupole components for the magnetic field of Her X-1 (a "butterfly" cross section), which allows accreting arclets on the NS surface. Determining the inner disc radius is not a trivial task in this case.
As a crude simplification, we estimate the magnetospheric radius for a purely quadrupolar field configuration based on the radius-dependence of the magnetic moment alone, taking $\mu_{\rm q}=B_{\rm q}R_*^4$ \citep[see e.g.][]{Barnard1982}. 

Compared to the dipole magnetic field, the quadrupole field decays more strongly with the radius, which has consequences for the inner edge of the accretion disc. To estimate the magnetospheric radius, we take an approach similar to the derivation of the Alfv\'en radius of a pure dipole field \citep[see p. 158 in][]{King2002}. First, we equate the magnetic pressure with the ram pressure of the gas at the magnetospheric radius $R_{\rm m,q}$:
\begin{eqnarray}
    P_{\rm mag}(R_{\rm m,q}) &=& P_{\rm ram}(R_{\rm m,q})\\
    \label{eq:presseq}
    \frac{\mu_{\rm q}}{8\pi R_{\rm m,q}^8} &=& \rho v^2\biggr\rvert_{R_{\rm m,q}},
\end{eqnarray}
for which we assume that the gas is in free fall and mass is conserved in spherical accretion, obeying 
\begin{eqnarray}
    \label{eq:v_ff}
    v_{\rm ff} &=&  \left(\frac{2GM}{r}\right)^{1/2}\\
    \label{eq:m_conserv}
    |\rho v| &=&  \frac{\dot{M}}{4\pi r^2}.
\end{eqnarray}
We substitute these expressions \eqref{eq:v_ff} and \eqref{eq:m_conserv} into equation \eqref{eq:presseq} and solve for $R_{\rm m,q}$:
\begin{eqnarray}
    \frac{\mu_{\rm q}}{8\pi R_{\rm m,q}^8} &=& \frac{(2GM)^{1/2}\dot{M}}{4\pi R_{\rm m,q}^{5/2}}\\
    \label{eq:r_m_derive}
    R_{\rm m,q} &=& \left( \frac{\mu_{\rm q}^{4}}{8GM\dot{M}^2} \right)^{1/11}.
\end{eqnarray}
Equation \eqref{eq:r_m_derive} can be expressed in unitless quantities and the quadrupole magnetic moment given in terms of the dipole moment $\mu_{\rm q}=\mu R_*$ for ease of comparison:
\begin{equation}
\label{eq:r_m_q}
    R_{\rm m,q} = 5.3\times 10^7 \Lambda_{\rm q} m^{1/11} \dot{M}_{16}^{-2/11} R_{*,6}^{4/11} \mu_{30}^{4/11}\ \rm{cm,}
\end{equation} 
introducing $\Lambda_{\rm q}$ to account for the accretion geometry. 
We note that $\Lambda_{\rm q}$ may be different from the $\Lambda$ for a dipole field. 

Using equation \eqref{eq:r_m_q}, we estimated the expected Keplerian frequency at the quadrupole-magnetospheric radius as
\begin{equation}
    f_{\rm K,q}= 3.6\ \Lambda_{\rm q}^{-3/2} m^{7/11} \dot{M}_{16}^{3/11} R_{*,6}^{-6/11} \mu_{30}^{-6/11} {\rm Hz.}
\end{equation}
It is evident that the quadrupole-magnetospheric radius is significantly smaller compared to the estimate for dipole field and
can thus explain how the accretion disc can get closer to the compact object and how the break frequencies consequently appear at higher frequencies. This possibility can be considered for both GRO~J1744$-$28 and Her~X-1. The quadrupole-magnetospheric radius with $\Lambda_{\rm q}\approx 0.6$ would be well in line with the break frequency observed for Her X-1\footnote{The small inner radius of Her X-1 poses a question of the critical fastness parameter since the source is close to torque equilibrium (exhibiting both spin-up and spin-down). We note that both the quadrupolar magnetic configuration and a tilted, warped disc would be likely to change the disc-magnetosphere interaction.}. 


In connection to the magnetic field strength discrepancy, as noted by \cite{2015MNRAS.454.2539M}, the CRSF of Her X-1 shows a positive luminosity-dependence up to $L\approx 4\times 10^{37}\,{\rm erg\,s^{-1}}$ \citep{2007A&A...465L..25S} at which a negative correlation is expected, that is, the critical luminosity should be higher than theoretically derived. However, the critical luminosity depends on the base length of the accretion channel on the NS surface. 
The presence of multipole components in the magnetic field of Her X-1 could lead to a complicated geometry of the accretion flow onto both the magnetic poles and arcs surrounding them \citep{Shakura1991,2013MNRAS.435.1147P}. 
The distribution of matter on a wider area would indeed allow Her X-1 to reach the observed luminosities without the formation of accretion columns according to the models of \citet{1976MNRAS.175..395B} and \citet{2015MNRAS.447.1847M}.\footnote{Based on \cite{2013MNRAS.435.1147P}, we take $4\times 10^9$~cm$^2$ to be a reasonable estimate for the hot spot area.} 
Consequently, we can infer that the presence of multipole fields in GRO~J1744$-$28 and Her~X-1 is plausible and can actually explain several of their peculiarities besides the unexpectedly high break frequencies. An intermediate polar known as V405~Aur may also have a complex magnetic field structure as it shows a large discrepancy between the magnetic field estimates based on polarization and corotation radius as noted by \cite{2019MNRAS.482.3622S}. However, since they could not measure the break frequency of V405~Aur, further work is required to confirm the scenario.

On the other hand, these objects appear to be exceptions highlighting the general rule which holds for the rest of our diverse sample of accreting NSs and white dwarfs. We conclude, therefore, that the break frequency can provide estimates for the magnetic field strength independent from the cyclotron line measurement and in a broader range of magnetic field strengths.
Even though a relatively large scatter still remains in this dependence, any significant deviations from it for sources not considered by us here shall be considered suspicious, i.e. may indicate presence of multipole field components or other unmodelled peculiarities (for instance, related to the accretion disc structure). The key to better accuracy lies in determining the timescale of initial variability in the accretion disc.

\section*{Acknowledgements}

This research was supported by the grant 14.W03.31.0021 of the Ministry of Science and Higher Education of the Russian Federation. 
The authors would like to acknowledge networking support by the COST Action GWverse CA16104 and Turku University Foundation. 
We also acknowledge the support from the Vilho, Yrj\"o and Kalle V\"ais\"al\"a Foundation, the Varsinais-Suomi Fund of the Finnish Cultural Foundation (JM), the Netherlands Organization for Scientific Research Veni Fellowship (AAM),  the German Research Foundation (DFG) grant WE 1312/53-1 (VFS), the Academy of Finland travel grants 333112, 331951, 349373, and 349906 (SST, JP), and the German Academic Exchange Service (DAAD) travel grant 57525212 (VD, VFS).

\section*{Data availability}

The data for X-ray pulsars is from the Heasarc \url{https://heasarc.gsfc.nasa.gov/cgi-bin/W3Browse/w3browse.pl}. Details of the reduced files and results presented in figures will be shared on reasonable request to the corresponding author. The pulse frequencies given in Table \ref{tab:pulsars} are from \textit{Fermi} Gamma-Ray Burst Monitor Pulsar Project at \url{https://gammaray.msfc.nasa.gov/gbm/science/pulsars.html}.

\bibliographystyle{mnras}
\bibliography{library}

\appendix

\section{Convolution}
\label{sec:convolution}

For many sources, the break frequency is situated amongst the harmonic frequencies. To confirm the presence of the break and measure its frequency more accurately, we took into account the modulation of the aperiodic noise by the periodic signal. We were especially motivated by the concerns of \cite{1997ApJ...476..267L} that the coupling of different variability sources widens the wings of regular pulsations and produces a plateau that is mistakable for a break in the power law. However, we decided not to use the analytical model presented in \cite{1997ApJ...476..267L} which would have had to include a large number of individual pulse harmonics to fit. Instead, a model power spectrum was constructed by numerically convolving two components: modelled broken power-law noise and the pulse profile power spectrum derived from data. 

In the time domain, we take the apparent X-ray energy flux $A(t)$ to be consistent of a noise component $N(t)$ modulated, i.e. multiplied, by the pulse signal $S(t)$.
This approximation is valid under the assumption that the beam pattern is independent on the mass accretion rate within the considered range. 
Both of these components can be separated into a constant and a zero-centred, time-dependent part: 
\begin{align}
    \begin{split}
        A(t) = N(t)S(t) &= [N_0+n(t)][S_0+s(t)] \\ &=  N_0S_0 + N_0s(t) +  S_0n(t) + n(t)s(t) .
    \end{split}
\end{align}

The Fourier transform turns the product in the time domain into a convolution in the frequency domain:
\begin{align}
    \begin{split} \label{eq:convo}
        \bar{A}(f) = \bar{N}(f)*\bar{S}(f) &= N_0S_0\delta(f) + N_0\bar{s}(f) \\ &+ S_0\bar{n}(f) + \bar{n}(f)*\bar{s}(f) \rm{,}   
    \end{split}
\end{align}
where the bar signifies the Fourier transform of the function and $*$ the convolution. 
Four terms in the right-hand side of (\ref{eq:convo}) correspond to $\delta$-function at $f=0$ (due to the average non-zero flux from the source), 
Fourier transform of the zero-centred pulsations,
broadband Fourier image of the mass accretion rate onto the NS surface, 
and Fourier image of pulsations disturbed by the fluctuations of the mass accretion rate, respectively. 
In general, the last term of (\ref{eq:convo}) results in more or less broad QPOs underlying the sharp peaks due to pulsations. 
These QPOs disturb the Fourier transform and broadband PDS.

We assumed the broken power-law noise describes the underlying continuum $\bar{n}(f)$ adequately, leaving out more complex models with the position of the break more difficult to infer. We adapted the \cite{1995A&A...300..707T} method for producing the power-law Fourier image for the case of a broken power law. The pulse profile was folded from the entirety of data to average out the aperiodic variability. The pulse profile was interpolated to light curve accuracy and repeated as a light curve to create a Fourier transform of only the regular pulsations, $\bar{s}(f)$.
Because it is not possible to determine the constants appearing in Eq.~\eqref{eq:convo} from observables, the final power spectrum was taken as a freely parametrised sum of the initial components $\bar{n}(f)$ and $\bar{s}(f)$, and their convolution $\bar{n}(f)*\bar{s}(f)$. The absolute value of the complex composition was calculated and the power level of Poisson noise was introduced as an additional free parameter for the fit. As a result, the model produces expected widened wings to the pulse peaks and gives the summed effect from the wings of all the pulse harmonics on the shape of the power spectrum close to the break frequency.

The convolved PDS was initially fit using \textsc{XSPEC} and Whittle statistics. The MCMC methods of \textsc{emcee} \citep{emcee} Python package were used to find the parameter uncertainties and inter-dependencies, starting from uniform priors for the parameters within relatively wide, yet feasible ranges. With 32 walkers, the number of iterations was 20000 with a burn-in of 1000.
See, for instance, \citet{2010MNRAS.402..307V} for discussion on how Whittle statistics is applied in the maximum likelihood estimation. 

\begin{figure}
\includegraphics[width=\textwidth]{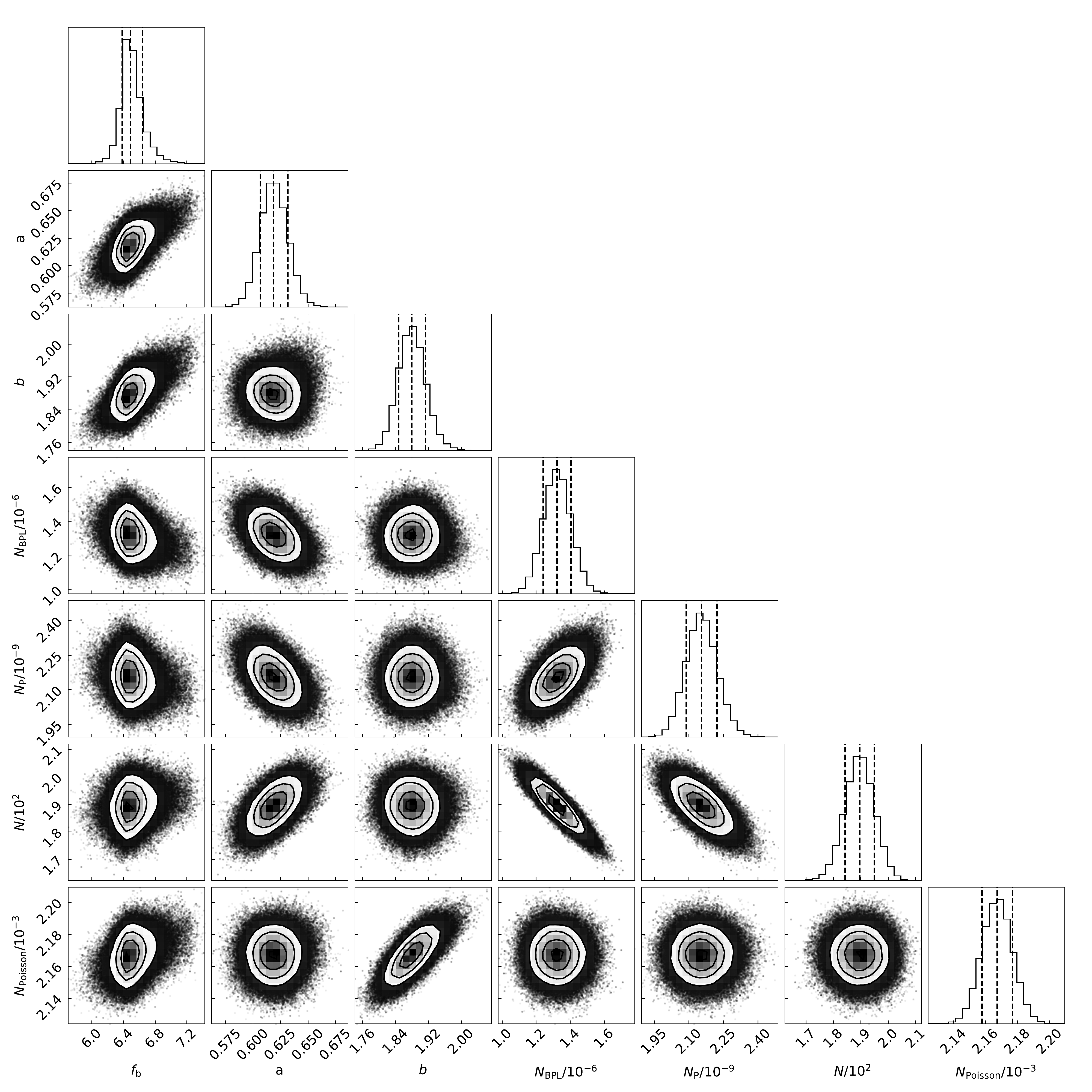}
\caption{Convolved power spectrum model for Her X-1: the distributions of individual parameters and their inter-dependencies as seen in the MCMC chain.}
\label{fig:emcee_herx1}
\end{figure}

The resulting parameter dependencies for Her~X-1 as discussed in the main text are presented in Fig.~\ref{fig:emcee_herx1}, plotted using the \textsc{corner} Python package \citep{corner}. Parameters are: the break frequency $f_{\rm b}$, the slope below break $a$, the slope above the break $b$, the normalisation of the broken power-law component $N_{\rm BPL}$, the normalisation of the PDS of regular pulsations $N_{\rm P}$, the normalisation of the sum $N$, and the level of Poisson noise $N_{\rm Poisson}$. 

\clearpage 


\bsp	
\label{lastpage}
\end{document}